\documentclass[prb,twocolumn,floatfix,showpacs]{revtex4}
\usepackage{graphicx,epsfig}
\usepackage{amsmath,amssymb,bm}
\begin{document}

\title{Influence of the contacts on the conductance of interacting
quantum wires}
\author{K.~Janzen}
\affiliation{Institut f\"ur Theoretische Physik, Universit\"at G\"ottingen, 
Friedrich-Hund-Platz 1, D-37077 G\"ottingen, Germany}
\author{V.~Meden}
\affiliation{Institut f\"ur Theoretische Physik, Universit\"at G\"ottingen, 
Friedrich-Hund-Platz 1, D-37077 G\"ottingen, Germany}
\author{K.~Sch\"onhammer}
\affiliation{Institut f\"ur Theoretische Physik, Universit\"at G\"ottingen, 
Friedrich-Hund-Platz 1, D-37077 G\"ottingen, Germany}

\begin{abstract}
We investigate how the conductance $G$ through a clean interacting quantum 
wire is affected by the presence of contacts and noninteracting
leads. The contacts are defined by a vanishing two-particle interaction
to the left and a finite repulsive interaction to the right or vice 
versa. No additional
single-particle scattering terms (impurities) are added. We first use
bosonization and the local Luttinger liquid picture and show that 
within this approach $G$ is determined by the properties of the 
leads regardless of the details of the spatial variation of the Luttinger 
liquid parameters. This generalizes earlier results obtained for 
step-like variations.
In particular, no single-particle backscattering 
is generated at the contacts.  We then study a microscopic model 
applying the functional renormalization group and show that the 
spatial variation of the interaction produces single-particle 
backscattering, which in turn leads to a reduced 
conductance. We investigate how the smoothness of the contacts affects 
$G$ and show that for decreasing energy scale its deviation
from the unitary limit follows a power law with the same exponent as 
obtained for a system with a weak single-particle impurity 
placed in the contact region of the interacting wire and the leads. 
\end{abstract}
\maketitle     

\section{Introduction}
\label{intro}

The low-energy physics of one-dimensional (1d) metals is not described by 
the Fermi liquid theory if two-particle interactions are taken into 
account. Such systems fall into the  Luttinger liquid (LL) universality 
class\cite{KS} that is characterized by power-law scaling of a variety of 
correlation functions and a vanishing quasi-particle weight. 
For spin-rotational invariant interactions and spinless models,
on which we focus here,  the exponents of the different correlation 
functions can be parametrized by a single number, the interaction dependent 
LL parameter $K < 1$ (for repulsive interactions; $K=1$ in 
the noninteracting case).  
As a second independent LL parameter one can take 
the velocity $v_c$ of charge excitations (for details see below).
Instead of being quasi-particles the low lying excitations 
of LLs are collective density excitations. This implies that 
impurities or more generally inhomogeneities have a dramatic effect 
on the physical properties of
LLs.\cite{LutherPeschel,Mattis,ApelRice,Giamarchi}

In the presence of only a single impurity on 
asymptotically small energy scales observables behave as if the  1d system 
was cut in two halfs at the position of the impurity, with open boundary 
conditions at the end points 
(open chain fixed point).\cite{KaneFisher,Furusaki0,Fendley} 
In particular, for 
a weak impurity and decreasing  energy scale $s$ the 
deviation of the linear conductance $G$ from the impurity-free  
value scales as $(s/s_0)^{2(K-1)}$, with $K$ being the scaling dimension 
of the perfect chain fixed point and $s_0$ a characteristic energy
scale (e.g.~the band width). This holds as long as 
$\left|V_{\rm back}/s_0\right|^2 (s/s_0)^{2(K-1)} \ll 1$, with 
$V_{\rm back}$ being a measure for the strength of 
the $2k_F$ backscattering of the impurity and $k_F$ the Fermi momentum.
For smaller energy scales or larger bare impurity backscattering this 
behavior crosses over to another power-law scaling 
$G(s) \sim (s/s_0)^{2(1/K-1)}$, with the scaling dimension of the open chain 
fixed point $1/K$. This scenario was verified for infinite 
LLs\cite{KaneFisher,Furusaki0,Fendley} as well as finite LLs  
connected to Fermi liquid leads,\cite{FurusakiNagaosa,Tilman} 
a setup that is closer to systems that can be realized in experiments. 
In the latter case the scaling holds as 
long as the contacts are modeled to be ``perfect'', that is free
of any bare and effective single-particle backscattering, and
the  impurity is placed in the bulk of the interacting quantum wire. 
For an impurity placed close to perfect contacts 
the exponents change to $2(K-1)/(K+1)$ (close to the perfect 
chain fixed point) and $1/K-1$ (close to the open chain fixed 
point).\cite{FurusakiNagaosa,Tilman} 

The role of an inhomogeneous two-particle interaction, that is an 
interaction that depends not only on the relative distance of the two 
particles, but also the center of mass is less well understood. 
In the present publication we will fill this gap.  
Such an inhomogeneity will generically appear close to the interface of 
the interacting quantum wire and the leads and a detailed understanding 
is thus essential for the interpretation of transport experiments
on quasi-1d quantum wires.\cite{experiments} 
We here use two models to study the effect of two-particle
inhomogeneities on the linear conductance. We first investigate the so-called
local Luttinger liquid (LLL) model,\cite{SS,MS,Ponomarenko} 
that is characterized 
by a spatial dependence of the LL parameters $K$ and $v_c$, with $K=1$ and 
$v_c=v_F$ in the leads ($v_F$ is the noninteracting Fermi velocity). 
We show that regardless of the details of the spatial variation the 
conductance always takes the perfect value $1/(2\pi)$ (in units such 
that $\hbar=1$ and the electron charge $e=1$). Thus the  LLL 
description cannot produce any effective 
single-particle backscattering from the contact region generated by 
an inhomogeneous two-particle interaction. Our results generalize 
earlier findings obtained for step-like variations of the LL 
parameters.\cite{SS,MS,Ponomarenko}  

We then study a microscopic lattice model with a spatially 
dependent nearest-neighbor 
interaction. Across the contact between the left lead and the wire 
the interaction is turned on from zero to a bulk value $U$ and 
correspondingly turned off close to the right contact. We show 
that this two-particle inhomogeneity generically leads to an effective
single-particle backscattering and a reduced conductance.  
To compute the latter we use an approximation scheme that is 
based on the functional renormalization group (fRG).\cite{Tilman}  
We numerically and analytically investigate the dependence of 
the conductance on  $U$, the length of the interacting 
wire $N$, and the ``smoothness'' with which the interaction is 
turned on and off. For weak effective inhomogeneities we analytically 
show that $1/(2 \pi) -G$ displays  scaling with the energy scale
$\delta_N=\pi v_F/N$ set by the length of the wire. The 
exponent we find is consistent with the one found for a weak
single-particle impurity close to a perfect 
contact.\cite{FurusakiNagaosa,Tilman} 
We give the energy scale up to which this scaling 
holds. It depends on the bulk interaction $U$ and the  
$2 k_F$ Fourier component of the function with 
which the interaction is varied close to the contacts. 
The latter provides a quantitative measure of the ``smoothness'' of
the contacts. This finding 
suggests a similarity between the universal low-energy properties of 
a LL with a single-particle inhomogeneity and a two-particle 
inhomogeneity. We discuss this similarity but also point out differences.

This paper is organized as follows. In Sec.~\ref{LLLsection} we study 
the two-particle inhomogeneity within the LLL picture. To motivate the
LLL model in Sec.~\ref{LLLmodel} we present a brief introduction into 
bosonization and the Tomonaga-Luttinger model.  The transport properties
of the LLL model are investigated in Sec.~\ref{chargeLLL}.
In Sec.~\ref{fRG} we introduce our microscopic model and give the basic 
equations of the fRG approach. We then discuss our numerical results 
in Sec.~\ref{subnum} and the analytical findings in Sec.~\ref{subana}. 
We close in Sec.~\ref{summary} with a summary.

\section{The local Luttinger liquid description}
\label{LLLsection}

In this section we discuss how sufficiently smooth contacts (perfect
contacts) can be properly described in a LLL picture. This approach 
was successfully used to show that for
perfect contacts the properties of the leads and not the finite 
size quantum wire determine the conductance.\cite{SS,MS,Ponomarenko} 
Up to now the LL parameters were always assumed to vary step-like. Here
we present a simple derivation of this result for an arbitrary
variation of the LL parameters. Also the role of impurities in the 
interacting wire was investigated within the LLL 
picture.\cite{Maslov2,FurusakiNagaosa} It was shown that the exponent
of the temperature dependence of the conductance is different for
impurities placed in the bulk and close to the
contacts.\cite{FurusakiNagaosa} Using the fRG this was later 
confirmed in a microscopic model in which the contacts were modeled 
to be arbitrarily smooth.\cite{Tilman}. 

\subsection{Model and generalized wave equations}
\label{LLLmodel}

In order to also elucidate the limitations of the LLL picture 
we first shortly recall the basic ideas necessary to justify the 
Tomonaga-Luttinger-model\cite{T,L,KS}
for interacting fermions in one dimension 
in the homogeneous case. We consider a system of length $L$ with
periodic boundary conditions. In second quantization the two-body
interaction reads
\begin{eqnarray}
\label{interaction}
  V & = &
\frac{1}{2}\int \int  v(x-x')
 \delta   \rho(x)\delta  \rho(x')  dx dx' \nonumber \\
&& +\frac{1}{2L}  {\cal N}^2 \tilde v(0)- \frac{1}{2}v(0)  {\cal N}, 
\end{eqnarray}
where $\delta   \rho (x) = \psi^\dagger (x)\psi (x)-   {\cal N}/L $
is the operator of the particle density 
relative to its homogeneous value, with $  {\cal N} $ the particle
 number operator. The Fourier transform of the two-body interaction
is denoted as $\tilde v(k)$. 
The last term is usually dropped as it only modifies
the chemical potential.
If the range of the two-body interaction is much
larger than the mean particle distance only particle-hole pairs in
the vicinity of the two Fermi points are present in the ground state
and the eigenstates with low excitation energy.\cite{T} This allows
to linearize the dispersion around the two Fermi points and to 
introduce two independent 
types of fermions, the right- and left-movers with particle
density operators $\delta   \rho_{\pm}(x)=   \rho_{\pm}(x)- {\cal N_\pm}$
\begin{eqnarray}
\label{density}
\delta  \rho_{\alpha}(x) & = & \frac{1}{L}\sum_{n\ne 0}e^{ik_nx}
  \rho_{n,\alpha}
 = \frac{\partial}{\partial x}\left [ \frac{-i}{2\pi}     
\sum_{n\ne 0} \frac{  e^{ik_nx}}{n}  \rho_{n,\alpha}
\right ] \nonumber \\ 
& \equiv & \frac{\partial  \Phi_{\alpha}}{\partial x}~. 
\end{eqnarray}
As shown below the field operators $ \Phi_{\alpha} $
are convenient objects for the solution
 of the problem.\cite{comment} 
In the subspace of low energy states the Fourier components
of the density $  \rho_{n,\alpha} $ obey the commutation relations\cite{T} 
\begin{equation}
\label{commutator}
[  \rho_{m,\alpha},  \rho_{n,\beta}] = \alpha m \, \delta_{\alpha
  \beta}\;\delta_{m,-n}~.
\end{equation}
After proper normalization they take the form of boson commutation relations. 

One can write the operator of the kinetic energy as a quadratic form of 
the   $  \rho_{\pm}$.\cite{T} The Tomonaga model then results from replacing
$\delta   \rho (x) $ by $\delta   \rho_+(x)+ \delta 
\rho_-(x) $
and $  {\cal N} $ by $  {\cal N}_++  {\cal N}_-  $ in 
Eq.~(\ref {interaction}). With the well known ``g-ology'' 
generalization\cite{KS}
the boson part of the interaction reads
\begin{eqnarray}
\label{VTomonaga}
&&   V_{\text{TL},b} =  \frac{1}{2} \int \!\! \int  \left\{ 
2g_2(x-x') \delta   \rho_+(x) \delta  \rho_-(x') 
 \right.  \\ && \left. + g_4(x-x') \left[
\delta   \rho_+(x)    \delta   \rho_+(x') + \delta 
\rho_-(x)\delta 
\rho_-(x') \right] \right\}dx dx' \; .\nonumber 
\end{eqnarray}
This reduces to Tomonaga's original model for $g_2\equiv g_4 \equiv
v$, while Luttinger\cite{L} later independently studied the model with
$g_4\equiv 0$. He also discussed the special case $g_2(x)\sim \delta
(x)$, which corresponds to the interaction term in the massless
Thirring model.\cite{TM} This simplifies various aspects but brings
in infinities which have to be removed e.g.~by normal ordering.
In the boson part of the kinetic energy
\begin{equation}
\label{kinE}
  T_b=\pi v_F \int\left[ \delta   \rho_+^2(x)
+ \delta   \rho_-^2(x) \right] dx
\end{equation}
the integrand has to be normal ordered. This, as above, is 
usually suppressed.

Inhomogeneous Luttinger liquids can be realized by adding an external
one-particle potential or by allowing the two-body interaction to
depend on both spatial variables, i.e.~$v(x-x')$ in 
Eq.~(\ref{interaction}) is replaced by $v(x,x')$. The operator of 
a one-particle potential can only be expressed in terms  of the 
$\delta   \rho_\alpha$ if it is sufficiently smooth in real space, 
i.e.~the external potential has a vanishing $2k_F$ Fourier component.
Similarly only for sufficiently smooth variations---in Sec.~\ref{subana}
we will specify the meaning of ``sufficiently smooth''---of $v$ 
with $(x+x')/2$ the analogous steps from Eq.~(\ref{interaction}) 
to Eq.~(\ref{VTomonaga}) i.e.~$g_i(x-x') \to g_i(x,x')$ are allowed 
without changing the low energy physics. The standard LLL model is 
obtained by assuming $g_i(x,x')=g_{i,x} \delta(x-x')$
\begin{widetext}
\begin{eqnarray}
\label{HLLL}
  H_{\text{LLL},b} &=& \pi \int \left \{ \left(v_F+\frac{g_{4,x}}{2\pi}\right)
 \left [ \delta   \rho_+^2(x)
+ \delta   \rho_-^2(x) \right ]+\frac{g_{2,x}}{\pi}
\delta   \rho_+(x) \delta 
\rho_-(x) \right \} dx \\
&=&
\frac{\pi}{2}  \int \left \{ \left(v_F+\frac{g_{4,x}+ g_{2,x} }{2\pi}\right)
 \left [ \frac{\partial\Phi^{(+)}(x)}{\partial x} \right ]^2
+\left(v_F+\frac{g_{4,x}- g_{2,x} }{2\pi}\right)
 \left [   \frac{\partial\Phi^{(-)}(x)}{\partial x}    \right ]^2
\right \}dx\\
 \nonumber
&\equiv&   
 \frac{\pi}{2}    \int \left \{v_N(x) \left[ 
\frac{\partial\Phi^{(+)}(x)}{\partial x}
\right]^2
+v_J(x) \left[ \frac{\partial\Phi^{(-)}(x)}{\partial x}
\right]^2   \right \} dx \nonumber~,
\end{eqnarray}
\end{widetext}
where we have defined $ \Phi^{(\pm)}(x)\equiv \Phi_+(x)
\pm \Phi_-(x) $, as well as the velocities $v_N(x)$ and $v_J(x)$.
As in the homogeneous model the right and left movers are coupled 
by the $g_2$ interaction and using the $\Phi^{(\pm)}(x) $
as the basic fields simplifies the solution as 
 $\Phi^{(\nu)}(x)$ and  $\Phi^{(\nu)}(x')$ commute. 
Apart from a term which vanishes in the thermodynamic limit
$\Phi^{(-\nu)}(x')$
 and $\partial \Phi^{(\nu)}(x)/\partial x $ obey canonical
commutation relations after proper normalization
\begin{equation}
\label{commutator2}
\left [ \frac{\partial\Phi^{(\nu)}(x) }{\partial x}
, \Phi^{(-\nu)}(x')  \right]= \frac{i}{\pi}\left [ 
\delta_L(x-x')-\frac{1}{L}\right] \; .
\end{equation}
This follows from  Eqs.~(\ref{density}) and (\ref{commutator}).
Here $\delta_L$ denotes the $L$-periodic delta function.
Neglecting the correction term yields the following Heisenberg 
equations of motion for the $\Phi^{(\nu)}(x,t)$
\begin{eqnarray}
\label{Heisenberg}
\frac{\partial}{\partial t} \Phi^{(+)}(x,t) & = &
-v_J(x)  \frac{\partial\Phi^{(-)}(x,t)}{\partial x} \; ,\nonumber \\
\frac{\partial}{\partial t} \Phi^{(-)}(x,t) & = &
-v_N(x)  \frac{\partial\Phi^{(+)}(x,t)}{\partial x} \;. 
\end{eqnarray} 
Therefore, the  $\Phi^{(\nu)}(x,t) $ obey generalized wave
equations, e.g.~for the field related to the change of the total density
\begin{equation}
\label{waveeq}
\frac{\partial^2}{\partial t^2} \Phi^{(+)}(x,t)
-v_J(x)\frac{\partial}{\partial x}v_N(x)
 \frac{\partial\Phi^{(+)}(x) }{\partial x}=0 \;.
\end{equation}
 The spatial derivative of the first equation in 
Eq.~(\ref{Heisenberg}) yields the continuity equation for the 
total charge density $\delta  \rho= \delta  \rho_++  
\delta  \rho_- $
\begin{equation}
\label{conteq}
\frac{\partial}{\partial t}\delta  \rho(x,t)
+\frac{\partial}{\partial x}\left[ v_J(x)
\frac{\partial\Phi^{(-)}(x,t)}{\partial x} \right]=0
\end{equation}
which implies the conservation of the total charge $Q=\int \delta \rho
dx$. As Eqs.~(\ref{waveeq}) and  (\ref{conteq}) are linear the
expectation values discussed in the following obey the same equations.

In a homogeneous system the velocities $v_J$ and $v_N$ are constant and the 
corresponding ``sound velocity'' called the charge velocity 
$v_c$ is given by $v_c=\sqrt{v_J v_N}$. It constitutes the first LL
parameter characterizing the system. The second LL parameter $K$ is 
defined as $K\equiv \sqrt{v_J/v_N}=v_c/v_N$. The general solution of 
the wave equation for constant $v_c$ and $K$ reads
$f(x-v_ct)+g(x+v_ct)$, where $f$ and $g$ are arbitrary functions.
 
\begin{figure}[tb]
\begin{center}
\includegraphics[width=0.45\textwidth,clip]{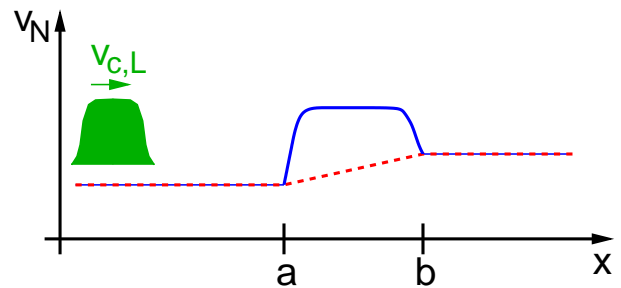}
\end{center}
\vspace{-0.6cm}
\caption[]{(Color online) Spatial dependence of 
velocity $v_N(x)$ used for the discussion
of the generalized wave equation (\ref{waveeq}). Two different
realizations (full and dashed curves) of the transition region 
$[a,b]$ are shown.  The shaded area shows the incoming density 
$\delta \rho (x,0)$.\label{fig1}}
\end{figure}

\subsection{Computing the transmitted charge}
\label{chargeLLL}

For general dependencies of $v_J(x)$ and $v_N(x)$ on the position the
generalized wave equation (\ref{waveeq}) cannot be solved
analytically. In the following we discuss properties of the solution 
in the thermodynamic limit when the velocities $v_J(x)$ and $v_N(x)$ 
are constant for $x<a$ and $x>b>a$ but have arbitrary 
(bounded) variation in the interval $[a,b]$.
Figure \ref{fig1} shows two qualitatively different types of behavior
of $v_N(x)$. The dashed curve presents a monotonous transition 
between the asymptotic values while the full curve 
corresponds to an ``interacting wire'' region in $[a,b]$
with a higher (constant)
value of $v_{N,w}$ corresponding to a stronger repulsive interaction.  
In order to describe the charge transport through this region we
consider an incoming density with compact support which at $t=0$
is completely to the left of $a$ and moves towards it with constant 
velocity $v_{c,L}$ (see Fig.~\ref{fig1}). Before the disturbance 
reaches point $a$ the
time evolution is
\begin{equation}
\label{shorttime}
\Phi^{(+)}(x,t)=f_{\rm in}(x-v_{c,L}t)\; , \;\;\; \delta\rho(x,t)=
f_{\rm in}'(x-v_{c,L}t) \; .
\end{equation}
The total incoming charge $Q_{\rm in}$ is given by $Q_{\rm in}=
f_{\rm in}(a)-f_{\rm in}(-\infty)$.
This short time solution for  $\Phi^{(+)}(x,t)$ determines
$\Phi^{(-)}(x,t)$ up to a constant. In the long time limit there will again
be no charge in the region $[a,b]$ with the total transmitted charge
$Q_{\rm trans}$  moving to the right with constant velocity $v_{c,R}$
and the total reflected charge $Q_{\rm ref}$  moving to the left with 
velocity $v_{c,L}$,  where charge conservation implies 
$Q_{\rm trans}+Q_{\rm ref}=Q_{\rm in} $. In the following we show that  
it is possible to determine $Q_{\rm ref}/ Q_{\rm in}$  without the
explicit solution for  $\delta\rho(x,t)$ in this long time limit and
without any additional assumptions on the spatial variation of
$v_N(x)$ and $v_J(x)$. The trick is to introduce the quantity
\begin{eqnarray}
\label{S(t)}
S(t) & \equiv &\int_{-\infty}^\infty \frac{1}{v_J(x)}\frac{\partial\Phi^{(+)}(x,t)}
{\partial t} dx\\
&=&  \Phi^{(-)}(-\infty,t)-\Phi^{(-)}(\infty,t)\nonumber
\end{eqnarray}
and to show that $S$ is time independent. 
This can be seen
by discussing the expression in the second line
which follows using Eq.~(\ref{Heisenberg}) or if one wants to
argue  with the density only by taking the time derivative and 
using the generalized wave equation
\begin{eqnarray}
\label{dS/dt}
\dot S(t)  &=&\int_{-\infty}^\infty 
\frac{1}{v_J(x)}\frac{\partial^2\Phi^{(+)}(x,t)} {\partial t^2} dx
\\ \nonumber & = & \int_{-\infty}^\infty
 \frac{\partial}{\partial x} \left[v_N(x) \frac{\partial}{\partial x} 
  \Phi^{(+)}(x) \right] 
dx \nonumber \\
&=&  v_{N,R} \delta \rho(\infty,t) - v_{N,L} \delta
\rho(-\infty,t)=0 \; ,\nonumber
\end{eqnarray}
where we have used that the assumption about the initial state
implies that $\rho(\pm\infty,t)$  
vanishes for all finite times.
The constant value of $S$ follows from Eq.~(\ref{shorttime})
\begin{eqnarray}
\label{S}
S & = & -\int_{-\infty}^a \frac{v_{c,L}}{v_{J,L}}f_{\rm
  in}'(x-v_{c,L}t)dx \nonumber \\ & = &
-\frac{v_{c,L}}{v_{J,L}}Q_{\rm in}=-\frac{Q_{\rm in}}{K_L} \; .
\end{eqnarray}
 When the incoming density hits the transition region 
$[a,b]$ it is partially reflected and transmitted,
where the detailed behavior is very different for the two
realizations of $v_N(x)$ shown 
in Fig.~\ref{fig1}. For sufficiently large times 
the total charge in the transition region
goes to zero and $ \Phi^{(+)}(x,t)$ takes the form
\begin{eqnarray}
\label{longtime}
\Phi^{(+)}(x,t)=\left \{ \begin{array}{ll}
 f_{\rm ref}(x+v_{c,L}t) \; , & {\rm for } \;  x<a \\
  \mbox{ const.}\; , &  {\rm for } \; a<x<b \\
 f_{\rm trans}(x-v_{c,R}t)\; , & {\rm for } \; x>b \; . \\
\end{array} \right.
\end{eqnarray} 
With the definition of $S$ in Eq.~(\ref{S(t)}) this yields
\begin{eqnarray}
\label{Slongtime}
S &=& \frac{v_{c,L}}{v_{J,L}} \int_{-\infty}^a
f_{\rm ref}'(x+v_{c,L}t)dx \nonumber \\ && 
- \frac{v_{c,R}}{v_{J,R}} \int_b^\infty
f_{\rm trans}'(x-v_{c,r}t)dx \nonumber \\
 &=&    \frac{Q_{\rm ref} }{K_L}-\frac{Q_{\rm trans}}{K_R} \; .
\end{eqnarray}  
The comparison of Eqs.~(\ref{S}) and (\ref{Slongtime}) 
as well as charge conservation leads to  the result derived 
earlier assuming a stepwise variation of the LL parameters\cite{SS,MS,Ponomarenko}
\begin{equation}
\label{ref,trans}
Q_{\rm ref}=\frac{K_L-K_R}{K_L+K_R}~Q_{\rm in} \; ,
\;\;\; Q_{\rm trans}=\frac{2K_R}{K_L+K_R}~Q_{\rm in} \; .
\end{equation}
Our result shows that the properties of the transition region 
play no role at all. The ratio $Q_{\rm trans}/Q_{\rm in}$ 
is positive, while $Q_{\rm ref}/Q_{\rm in} $ has no definite 
sign.

From this result one can easily obtain the linear conductance $G$
through the system. We first discuss a homogeneous system with
constant LL parameters $K_L$ and $v_{c,L}$ (the use of the index $L$
becomes clear later).  We consider a current free 
initial state in which the density is increased by 
$\delta \rho_0$ in the left half of the infinite system. Then 
half of the additional density moves to the right with 
velocity $v_{c,L}$ which corresponds to the current
\begin{equation} 
j=\frac{1}{2}v_{c,L}\delta \rho_0=\frac{1}{2}v_{c,L}\frac{\partial
  \rho}{\partial \mu}~\delta \mu_0 \equiv G ~\delta \mu_0
\end{equation}
in the central region extending linearly with time.
Here $\mu$ denotes the chemical potential. 
With $\partial \rho/\partial \mu=1/(\pi v_N)$ this yields for the
homogeneous system
\begin{equation} 
 G_{\rm hom}=\frac{K_L}{2\pi} \; .
\end{equation}
We now switch to an inhomogeneous system as in Fig.~\ref{fig1}.
In analogy to the initial condition in the homogeneous case we raise 
the density for $x < a$ by $\delta \rho_0$. 
The stationary current is then obtained by multiplying the result 
for the homogeneous case by the fraction of transmitted charge 
through $[a,b]$, computed in Eq.~(\ref{ref,trans}). For the
conductance this leads to 
\begin{equation} 
 G_{\rm inhom}=\frac{2K_R K_L}{K_L+K_R}~\frac{1}{2\pi} \; ,
\end{equation}
which again is independent of the details of the transition region.

For an interacting quantum wire in region $[a,b]$ attached to  
noninteracting leads ($K_L=K_R=1$) the conductance is 
$1/(2\pi)$ independently of how quickly the LL parameters vary 
spatially near the contact points $a$ and $b$. This shows that 
the  LLL description cannot produce one-particle backscattering 
from the contact region generated by an inhomogeneous two-particle  
interaction. As we will discuss next using a microscopic model and the
fRG approach contacts defined by a vanishing interaction to the left
and a positive finite interaction to the right (or vice versa)
generically produce an effective single-particle backscattering. 
This clearly shows the limitations of the LLL picture.
With the approximation to describe the two-body interaction 
as a quadratic form in the bose-fields the dependence of the
conductance on the sharpness of the transition is lost. 
In contrast, our microscopic model directly allows to study the
transition from smooth to abrupt contacts with the concomitant change
of the conductance.  
    
\section{A microscopic model}
\label{fRG}

As our microscopic model we consider the spinless tight-binding
Hamiltonian with nearest-neighbor hopping $\tau>0$ and a spatially dependent
nearest-neighbor interaction $U_{j,j+1} = U_{j+1,j}$
\begin{eqnarray}
\label{ham}
H & = & -\tau \sum_{j=-\infty}^{\infty} \left( \,
 c^{\dag}_{j+1} c_j^{\phantom\dag} + c^{\dag}_j \, c_{j+1}^{\phantom\dag}
 \, \right) \nonumber \\
&&+ \sum_{j=1}^{N-1} U_{j,j+1}  \left( n_j - 1/2
\right)  \left( n_{j+1} - 1/2 \right) \; ,
\end{eqnarray}
where we used standard second-quantized notation with $c^{\dag}_j$ 
and $c_j^{\phantom\dag}$ being creation and annihilation operators 
on site $j$, respectively, and $n_j = c^{\dag}_j \,
c_j^{\phantom\dag}$ the local density operator. 
The interaction acts only between the bonds of the sites 
$1$ to $N$, which define the interacting wire. We later take
\begin{eqnarray}
\label{hdefdef}
U_{j,j+1} = U h(j) \, , \;\;\; j=1,2,\ldots,N-1 \; , 
\end{eqnarray}
with a function $h$ that is different from 1
only in regions close to the contacts at sites 1 and $N$. 
We here mainly consider the half-filled band case $n=1/2$. 
In the interacting region the fermions have higher energy compared to
the leads. 
To avoid that the interacting wire is depleted (implying a vanishing
conductance) we added a compensating single-particle 
term. It can be included as a  shift of the local density operator by
$1/2$. This ensures that the average filling in the leads and the wire
is $n=1/2$. It is important to have a definite filling in the
interacting wire as the LL parameters depend on $n$. For general
fillings the required shift of the density 
depends on $n$ and the interaction.\cite{Tilman}

The model Eq.~(\ref{ham}) with nearest neighbor interaction $U$ 
across all bonds (not only the once within $1$ and $N$) is a LL for
all $U$ and fillings, except at half filling for $|U| >
2\tau$.\cite{Haldane} At $n=1/2$ the LL parameter 
$K$ is given by\cite{Haldane} 
\begin{equation}
\label{BetheAnsatz}
 K = \left[\frac{2}{\pi} \, 
 \arccos \left(-\frac{U}{2\tau} \right) \right]^{-1} \; ,
\end{equation}
for $|U| \leq 2 \tau$. To leading order in $U/\tau$ this gives
\begin{equation}
\label{leadingorder}
 K = 1- \frac{U}{\pi \tau} + {\mathcal O}\left([U/\tau]^2\right) \; .
\end{equation}
 
At temperature $T=0$ the linear conductance of the system 
described by Eq.~(\ref{ham}) can be written as\cite{footnote1} 
\begin{eqnarray}
G(N) = \frac{1}{2 \pi} \; |t(0,N)|^2  
\label{conductf}
\end{eqnarray} 
with the effective transmission $|t(\varepsilon,N)|^2 = 
(4 \tau^2- [\varepsilon+\mu]^2) |{\mathcal G}_{1,N}(\varepsilon,N)|^2$.
The one-particle Green function 
${\mathcal G}$ has to be computed in the presence of interaction 
and in contrast to the noninteracting case acquires an $N$ 
dependence. To compute ${\mathcal G}$ and thus $G$ 
we use a recently developed fRG scheme.\cite{VM1} The starting 
point is an exact hierarchy of differential flow equations for the 
self-energy $\Sigma^{\Lambda}$ and higher
order vertex functions, where $\Lambda \in (\infty,0]$ 
denotes an infrared energy cutoff which is the flow parameter.
We here introduce $\Lambda$ as a cutoff for the Matsubara frequency.
A detailed account of the method is given in
Ref.~\onlinecite{Tilman}. 
We truncate the hierarchy by neglecting the flow of the two-particle
vertex only considering $\Sigma^{\Lambda}$, which is then
energy independent. The self-energy $\Sigma^{\Lambda=0}$ at the end of
the fRG flow provides an approximation for the exact $\Sigma$. 
This approximation scheme and variants of it  were successfully 
used to study a variety of transport problems through 1d  wires 
of correlated electrons. 
In particular, in all cases of inhomogeneous LLs studied the 
exponents of power-law scaling were reproduced correctly to leading
order in $U$.\cite{VM2,Tilman,Xavier,Xavier2}

On the present level of approximation the fRG flow equation for the
self-energy reads
\begin{equation}
 \frac{\partial}{\partial\Lambda} \Sigma^{\Lambda}_{1',1} =
 - \frac{1}{2\pi} \sum_{\omega = \pm \Lambda} \sum_{2,2'} \,
 e^{i\omega 0^+} \, {\mathcal G}^{\Lambda}_{2,2'}(i\omega) \,
 \Gamma_{1',2';1,2} \; ,
\label{flowsigma}
\end{equation}
where the lower indices $1$, $2$, etc.\ label single-particle states,
$\Gamma_{1',2';1,2}$ is the bare antisymmetrized two-particle
interaction and
\begin{equation}
 {\mathcal G}^{\Lambda }(i\omega) = 
 \left[ {\mathcal G}_0^{-1}(i\omega) - \Sigma^{\Lambda} \right]^{-1} \; ,
\label{Gdef}
\end{equation}
with the noninteracting propagator 
${\mathcal G}_0$. For the initial condition of the self-energy 
flow see below. 
As our single-particle basis
we later use Wannier states  [as in the Hamiltonian (\ref{ham})] 
as well as momentum states. 

\subsection{Numerical results}
\label{subnum}

In the real space Wannier basis the self-energy matrix 
is tridiagonal and 
the set of coupled differential equations (\ref{flowsigma}) 
reads (with $j \in [1,N]$)
\begin{eqnarray}
\frac{d}{d \Lambda}\Sigma^{\Lambda}_{j,j} = - \frac{1}{2\pi} 
\sum_{l=\pm 1} \sum_{\omega = \pm \Lambda} U_{j,j+l} 
{\mathcal G}^{\Lambda}_{j+l,j+l} (i \omega) \,
 e^{i\omega 0^+} \; ,
  \label{diffsystem1} \\
\frac{d}{d \Lambda} \Sigma^{\Lambda}_{j,j\pm 1} =  \frac{U_{j,j\pm
    1}}{2\pi} 
  \sum_{\omega = \pm \Lambda}
  {\mathcal G}^{\Lambda}_{j,j \pm 1} (i \omega) \,
 e^{i\omega 0^+} \, .
\label{diffsystem}
\end{eqnarray}  
To derive these equations we
have used that
\begin{equation}
\label{vertex}
 \Gamma_{j'_1,j'_2;j_1,j_2} =
\bar U_{j_1,j_2} \, ( \delta_{j_1,j'_1} \delta_{j_2,j'_2} - 
   \delta_{j_1,j'_2} \delta_{j_2,j'_1} )
\end{equation}
with $\bar U_{j_1,j_2} = U_{j_1,j_1+1} 
(\delta_{j_1,j_2-1} + \delta_{j_1,j_2+1})$.
The initial condition 
of $\Sigma^\Lambda$ at $\Lambda=\infty$ is\cite{VM1,Tilman} 
 \begin{eqnarray}
\label{initialcondition}
\Sigma^{\infty}_{j,j} & = & -\left( U_{j-1,j} +
  U_{j,j+1} \right)/2 \; ,\nonumber \\  
\Sigma^{\infty}_{j,j\pm 1} & = & 0 \; .
\end{eqnarray}
Even for very large $N$ (earlier results were obtained for up to
$N=10^7$)\cite{Tilman,VM2} the system Eqs.~(\ref{diffsystem1}) 
and (\ref{diffsystem}) can easily be integrated numerically starting 
at a large but finite $\Lambda_0$. Due to the slow decay of the 
right-hand-side (rhs) of the flow equation the integration from 
$\infty$ to $\Lambda_0$ gives a nonvanishing contribution on the 
diagonal of the self-energy matrix that does not vanish even for 
$\Lambda_0 \to \infty$ and that exactly cancels the term in 
Eq.~(\ref{initialcondition}) such that
 \begin{eqnarray}
\label{initialcondition1}
\Sigma^{\Lambda_0}_{j,j} & = & 0 \; ,\nonumber \\  
\Sigma^{\Lambda_0}_{j,j\pm 1} & = & 0 \; .
\end{eqnarray}

For half filling the Hamiltonian (\ref{ham}) is particle-hole
symmetric. For $N$ odd and symmetric $h(j)$, that is $h(j)=h(N-1-j)$,
on which we focus here, it is furthermore invariant under inversion at site
$j=(N+1)/2$. Together these symmetries lead to a vanishing 
conductance as was discussed earlier.\cite{Oguri1,VM3} We thus only
consider even $N$. The two extreme cases of (a) an abrupt turning on and
off of the interaction $h(j)=1$ for $j=1,2,\ldots,N-1$, and zero
otherwise, and (b) a very smooth variation of $h(j)$ 
were considered earlier. In case (a) $1/(2\pi)-G$ increases for 
fixed $N$ and increasing\cite{VM2,VM3} $U$ as well as for fixed 
$U$ and increasing $N$.\cite{VM2} Using our fRG scheme
it was shown that for asymptotically large $N$, $G$ vanishes as 
\begin{eqnarray}
\label{vanish}
G \sim (\delta_N/\tau)^{2(1/K-1)} \; ,
\end{eqnarray}
with the energy scale $\delta_N=\pi v_F/N$. The scale on
which the asymptotic low-energy scaling sets in strongly depends on $U$
and even for up to $N=10^6$ it was only reached for fairly large $U$,
e.g.~$U=1.5 \tau$. 
As discussed in the introduction scaling of $G$ with the same exponent 
as in Eq.~(\ref{vanish}) is found for a single impurity in an
otherwise perfect chain.
This suggests that the low-energy physics 
of a correlated quantum wire with two contact inhomogeneities due 
to the two-particle interaction is similar to that of a wire with a
single impurity (single-particle term). Here we will further 
investigate this relation. 

Already at this stage it is important to note that the similarity 
is not complete. As shown elsewhere\cite{Severin} $G(T)$ 
for two contacts, fixed $U$, fixed large $N$, and 
decreasing temperature $T$  vanishes as (for $T \gg \delta_N$) 
\begin{eqnarray}
\label{vanish2}
G \sim (T/\tau)^{1/K-1} \; ,
\end{eqnarray}
that is with an exponent that is only one-half of the one found in the 
$\delta_N$ scaling Eq.~(\ref{vanish}). This is in clear contrast 
to the temperature dependence of $G$ 
for a single impurity in an otherwise perfect chain. In this case 
the exponent $2(1/K-1)$ is found in the scaling with any infrared 
energy scale, e.g.~$T$, $\delta_N$, and $\Lambda$. 

For a given $N$ and $U$ it is always possible to find a sufficiently 
smooth function $h(j)$ 
such that $1/(2\pi) -G$ is smaller then an arbitrarily small 
upper bound [case (b)].\cite{Tilman,VM2} 
In that case the contacts are perfect and the effect of impurities in 
the interacting wire connected to Fermi liquid leads can be studied.
Further down we will give a quantitative definition of the meaning of 
``sufficiently smooth''.

\begin{figure}[tb]
\begin{center}
\includegraphics[width=0.45\textwidth,clip]{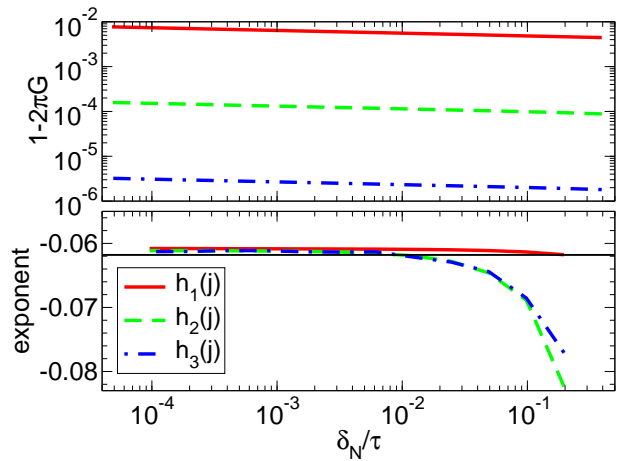}
\end{center}
\vspace{-0.6cm}
\caption[]{(Color online) {\it Upper panel:} The conductance $1- 2\pi
  G$ as a function of $\delta_N/\tau$ for $U/\tau=0.2$, contacts of $m=7$ lattice
  sites, and three different contact functions $h_i(j)$ of 
increasing smoothness. 
{\it Lower panel:} The effective exponent (logarithmic derivative) of
the data. The thin solid line shows the exponent $2(K-1)/(K+1)$ with
$K(U/\tau=0.2) = 0.9401$.  
\label{fig2}}
\end{figure}

Here we study the dependence of the conductance on $N$, $U$, and the
function $h(j)$ by numerically solving the flow equations 
(\ref{diffsystem1}) and (\ref{diffsystem}) with the initial 
condition Eq.~(\ref{initialcondition1}). 
In Fig.~\ref{fig2} we show $1-2 \pi G$ as a function of 
$\delta_N/\tau\sim 1/N$ on a log-log scale (upper panel) for $U/\tau=0.2$ and three 
different $h(j)$ of increasing smoothness.   
In our numerical calculations 
for simplicity we focus on contact  functions 
$h(j)$ that are symmetric 
around the middle of the interacting wire (equal contacts) and 
therefore only give their definition for the first half of the wire 
(with $j \in [1,N-1]$)
\begin{eqnarray*}
\label{h1def}
h_1(j) & = &  1  \, , \;\;  j=1,\ldots,N/2 \; , \\
\label{h2def}
h_2(j)  & = &  \left\{  \begin{array}{ll}
 j/m  \, , & \;\;  j=1,\ldots, m-1  \\
1  \, , & \;\;  j=m,\ldots, N/2  \end{array}
\right. \; ,
\end{eqnarray*}
and 
\begin{eqnarray*}
\label{h3def}
h_3(j) =   \left\{  \begin{array}{ll}  
    2(j/m)^2  \, , & \; j=1,\ldots, (m-1)/2  \\
   1-2(j/m-1)^2 \, , & \; j=(m+1)/2,\ldots,m-1 \\
1  \, , &  \; j=m,\ldots, N/2
  \end{array} \right. ,
\end{eqnarray*}
where $m$ is odd and measures the length of the contact region. 
In Fig.~\ref{fig2} we chose $m=7$. 
The lower panel of Fig.~\ref{fig2} shows the logarithmic derivative of
the data. A plateau in this figure corresponds to power-law scaling of
the original data with an exponent given by the plateau value. 
For fixed $\delta_N$, $1- 2 \pi G$ decreases quickly with increasing
smoothness of  $h(j)$.  
For sufficiently small $\delta_N/\tau$  all curves follow power-law scaling. 
The smoother the contact the smaller $\delta_N/\tau$ has to be before the
scaling sets in. 
The numerical exponent is close to $2(K-1)/(K+1)$ (shown as the thin
solid line), with $K$ from Eq.~(\ref{BetheAnsatz}). The latter is 
the exponent found for a system with perfect
contacts and a weak single-particle impurity placed in the contact
region. We also studied other values for the length of contact 
$m$ and found similar results. The deviation $1- 2 \pi G$ 
from the unitary limit decreases if $m$ is increased while 
all other parameters are fixed. 

Within our approximation scheme we map the many-body problem on an
effective single-particle problem with the energy independent
$\Sigma^{\Lambda=0}$ as an impurity potential.\cite{VM1,Tilman} 
During the fRG flow the interplay of the two-particle 
inhomogeneity and the bulk interaction generates an oscillating 
self-energy with an amplitude that decays slowly away from the 
two contacts. 
This is in close analogy to 
the case of a single impurity in an otherwise perfect 
chain.\cite{Glazman,VM1,Tilman} Scattering off this effective 
potential leads to the reduced conductance.  

\begin{figure}[tb]
\begin{center}
\includegraphics[width=0.45\textwidth,clip]{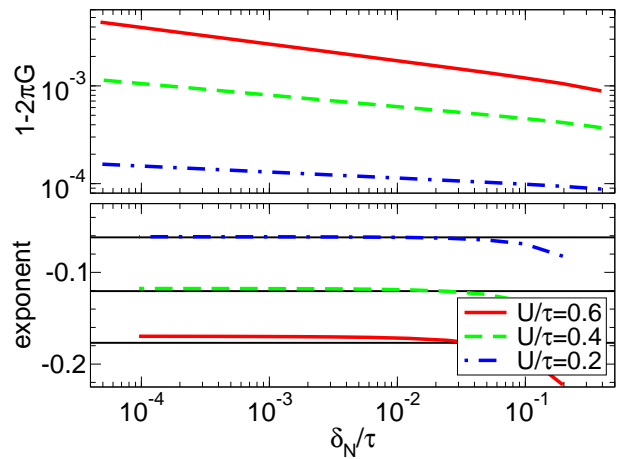}
\end{center}
\vspace{-0.6cm}
\caption[]{(Color online) {\it Upper panel:} The conductance $1- 2\pi
  G$ as a function of $\delta_N/\tau$ for the contact function 
 $h_2(j)$,  contacts of $m=7$ lattice
  sites, and $U/\tau=0.2$, $0.4$, $0.6$. 
{\it Lower panel:} The effective exponent (logarithmic derivative) of
the data. The thin solid lines show the exponents $2(K-1)/(K+1)$ with
$K(U/\tau=0.2) = 0.9401$, $K(U/\tau=0.4) = 0.8864$, and $K(U/\tau=0.6) = 0.8375$.  
\label{fig3}}
\end{figure}

The $U$ dependence of $1-2 \pi G$ and the effective exponent is
shown in Fig.~\ref{fig3} for $h_2(j)$. At fixed $\delta_N$ the deviation of
the conductance from the unitary limit increases with $U$. 
For all $U/\tau$ shown we find power-law behavior with exponents that are
close to $2(K-1)/(K+1)$ (again shown as the thin solid lines).
The larger $U/\tau$ the larger is the deviation of the numerical
exponent and $2(K-1)/(K+1)$. This is not surprising as our approximate
scheme can only be expected to capture exponents to leading order in
$U/\tau$.\cite{Tilman}

\begin{figure}[tb]
\begin{center}
\includegraphics[width=0.45\textwidth,clip]{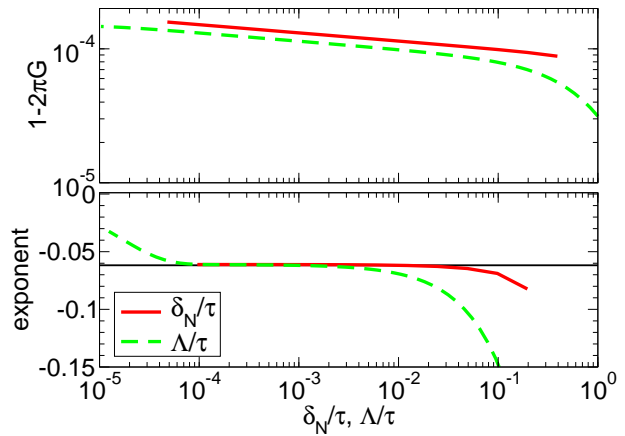}
\end{center}
\vspace{-0.6cm}
\caption[]{(Color online) {\it Upper panel:} The conductance $1- 2\pi
  G$ as a function of $\delta_N/\tau$ and $\Lambda/\tau$ (with $N=2^{16}$ sites)
for the contact function $h_2(j)$,  contacts of $m=7$ lattice
sites, and $U/\tau=0.2$.  
{\it Lower panel:} The effective exponent (logarithmic derivative) of
the data. The thin solid line shows the exponents $2(K-1)/(K+1)$ with
$K(U/\tau=0.2) = 0.9401$.  
\label{fig4}}
\end{figure}

For later reference we note that power-law behavior is not only found
in the $\delta_N$ (that is $1/N$) dependence of $1 - 2\pi G$ but 
also in the dependence on the fRG flow parameter $\Lambda$ at a large
fixed $N$. In this case $G$ is computed using Eq.~(\ref{conductf})
with the Green function at scale $\Lambda$. This is shown in
Fig.~\ref{fig4} where we compare the dependence on $\delta_N/\tau$ and 
$\Lambda/\tau$ for $U=0.2$ and $h_2(j)$. Scaling with $\Lambda/\tau$  holds
roughly down to the scale $\delta_N/\tau=\pi v_F/(\tau N)$  set by the length
of the interacting wire. Beyond this scale $G$ saturates. 
   
We now analytically confirm the power-law scaling of $1/(2 \pi) - G$
for weak effective inhomogeneities and compute the exponent to
leading order in the bulk interaction.
 
\subsection{Analytical results}
\label{subana}

Numerically we found that $1/(2 \pi)-G$ 
shows power-law scaling in both effective low-energy cutoffs $\delta_N$ and 
$\Lambda$, as long as the scaling variable considered is sufficiently 
larger than the other fixed energy scale. We note that the same holds for
the temperature $T$ as the variable.\cite{Sabineprivate} In all cases
the same exponent is found. Analytically the easiest access to scaling
is by considering $N \to \infty$, $T=0$, and taking $\Lambda$ as the
variable. We will here follow this route. 
For the case of a single impurity with bare backscattering amplitude
$V_{\rm back}$ in an infinite LL (no Fermi liquid leads) a similar
calculation was performed within the fRG\cite{VM1} 
to show that for small $\Lambda$  
\begin{eqnarray}
\Sigma_{k_F,-k_F}/\tau \sim (V_{\rm back}/\tau) \; 
(\Lambda/\tau)^{-U/(\pi \tau)} \; .
\label{bulkstuff}
\end{eqnarray}
Within the Born approximation this leads to 
\begin{eqnarray}
\frac{K}{2 \pi} - G \sim \left|V_{\rm back}/\tau\right|^2 
(\Lambda/\tau)^{-2 U/(\pi
\tau)} \; ,
\label{bulkstuff2}
\end{eqnarray}
where $-2 U/(\pi \tau)$ is the leading order approximation of the weak
impurity exponent $2(K-1)$ [see Eq.~(\ref{leadingorder})].
This scaling holds as long as the rhs stays small. 
For analytical calculations it is 
advantageous to switch from the real space basis to the momentum
states. 

In the momentum state basis the flow equation (\ref{flowsigma}) is
given by 
\begin{equation}
 \frac{\partial}{\partial\Lambda} \Sigma^{\Lambda}_{k',k} =
 - \frac{1}{2\pi} \sum_{\omega = \pm \Lambda}  \int_{-\pi}^{\pi} dq \,
 dq' \, e^{i\omega 0^+} \, {\mathcal G}^{\Lambda}_{q,q'}(i\omega) \,
 \Gamma_{k',q';k,q} \; ,
\label{flowsigmamomentum}
\end{equation}
with
\begin{eqnarray}
 \Gamma_{k',q';k,q} & = & \frac{1}{2 \pi}\left[ e^{i(q-q')} -
   e^{i(q-k')} + e^{i(k-k')} - e^{i(k-q')} \right]  \nonumber \\
 && \times \tilde U(k+q-k'-q') 
\label{vertexmomentum}
\end{eqnarray}
and the Fourier transform of the interaction 
\begin{eqnarray}
\tilde U(k)  & = & \frac{1}{2 \pi} 
\sum_{j=1}^{N-1} U_{j+1,j} e^{i j k} \nonumber \\
 & = & U \; \tilde h(k)  \; .
\label{FTU}
\end{eqnarray}
Here $\tilde h(k)$ denotes the Fourier transform of the function 
$h(j)$ that contains the shape of the turning on and off of the 
interaction. In the present section we do not assume special symmetry 
properties of $h(j)$. Thus the two contacts might be different, 
as it is generically the case in experiments. 
At $\Lambda=\Lambda_0$ all matrix elements of $\Sigma^{\Lambda_0}$ are
zero. We now expand the rhs of Eq.~(\ref{flowsigmamomentum}) to first 
order in $\Sigma^{\Lambda}$. This expansion is justified as long as
$\Sigma^{\Lambda}$ remains small (it is certainly small at the
beginning of the fRG flow). The flow equation then becomes an
inhomogeneous, linear differential equation 
\begin{eqnarray}
 \frac{\partial}{\partial\Lambda} \Sigma^{\Lambda}_{k',k} 
= {\mathcal F}^{(1)}_{k',k}(\Lambda) + {\mathcal F}^{(2)}_{k',k}
\left(\Lambda,\Sigma^{\Lambda}\right) \; ,
\label{dglstructure}
\end{eqnarray} 
with 
\begin{eqnarray}
{\mathcal F}^{(1)}_{k',k}(\Lambda) =  - \frac{1}{2\pi} \sum_{\omega = \pm
  \Lambda}  \int_{-\pi}^{\pi} dq \,
\frac{\Gamma_{k',q;k,q} }{i \omega - \xi_q}  \nonumber \\
=  -\frac{U}{2 \pi \tau} \tilde h(k-k')  \left( e^{ik} +
  e^{-ik'}\right)   
\left(1 - \frac{\Lambda}{
  \sqrt{\Lambda^2 + 4 \tau^2}}\right)
  \, ,
\label{F_1def}
\end{eqnarray} 
where we used $\xi_k = -2 \tau \cos{(k)} - \mu$ and Eq.~(\ref{vertexmomentum}).
The factor $e^{i\omega 0^+}$ was dropped as the integration starts at
$\Lambda_0 < \infty$. Later we will primarily be interested in the
$k_F,-k_F$ matrix element (backscattering) of $\Sigma$. For these
momenta the rhs simplifies to
\begin{eqnarray}
{\mathcal F}^{(1)}_{k_F,-k_F}(\Lambda)
=  -\frac{U}{\pi \tau} \tilde h(-2k_F) e^{-i k_F}
\left(1 - \frac{\Lambda}{
  \sqrt{\Lambda^2 + 4 \tau^2}}\right)
\; .
\label{F_1defkF}
\end{eqnarray} 
The term on the rhs of the linearized flow equation that contains
$\Sigma^\Lambda$ is 
\begin{eqnarray}
{\mathcal F}^{(2)}_{k',k}(\Lambda,\Sigma^\Lambda)  & = &
- \frac{1}{2\pi}   
\sum_{\omega = \pm \Lambda}  \int_{-\pi}^{\pi} dq \,
dq' \, \Gamma_{k',q';k,q} \nonumber \\ && \times 
\left({\mathcal G}_0 \Sigma^{\Lambda} 
{\mathcal G}_0 \right)_{q,q'} \; .
\label{F_2def}
\end{eqnarray} 
Before further evaluating this term we take $N \to
\infty$ and compute $ \Gamma_{k',q;k,q}$ in this limit. As we will see
in this case the information about the smoothness of $h(j)$ is
lost on the rhs of Eq.~(\ref{F_2def}). 
It nevertheless enters the solution of the differential equation 
via the inhomogeneity ${\mathcal F}^{(1)}_{k',k}$ in
Eq.~(\ref{dglstructure}). As we are only  interested in 
the leading divergent behavior of the linear part on the rhs of 
Eq.~(\ref{dglstructure}) this procedure is justified.
To be specific we first assume that the interaction is turned on and
off abruptly, that is $h(j) =1$ for $j=1,2,\ldots,N-1$ and zero
otherwise. To obtain the Fourier transform of the two-particle
interaction we have to perform
\begin{eqnarray}
\label{halbdelta}
\tilde U (k) & = & U \tilde h(k) = \frac{U}{2 \pi} \sum_{j=1}^{N-1}
e^{ijk} \nonumber \\
& = & \frac{U}{2 \pi} \frac{e^{ik}-e^{iNk}}{1-e^{ik}}
\nonumber \\
&  \stackrel{N \to \infty}{\longrightarrow}& U \, \frac{1}{2} \,
 \delta_{2\pi}(k) \; .
\end{eqnarray}
With this result the two-particle vertex simplifies to
\begin{eqnarray}
\label{vertexsimple}
\Gamma_{k',q;k,q} & \stackrel{N \to \infty}{\longrightarrow} &
\frac{U}{\pi} \;  
\left[ \cos(k-k') - \cos(q-k') \right]  \nonumber \\
&& \times \delta_{2\pi}(k+q-k'-q') \; .
\end{eqnarray}
Up to a factor $1/2$ the vertex is then equivalent to the one obtained
for a homogeneous nearest neighbor interaction. 
With this vertex ${\mathcal F}^{(2)}_{k',k}(\Lambda,\Sigma^\Lambda)$
Eq.~(\ref{F_2def}) reads
\begin{eqnarray}
{\mathcal F}^{(2)}_{k',k}(\Lambda,\Sigma^\Lambda) =  
- \frac{U}{2\pi^2}   \sum_{\omega = \pm \Lambda}  
\int_{-\pi}^{\pi}  dq [ \cos(k-k') 
 \nonumber \\  - \cos(q-k')] 
\frac{\Sigma_{q+k-k',q}}{(i \omega - \xi_{q+k-k'})(i \omega -
  \xi_q)} \; .
\label{F2eval}
\end{eqnarray}

The differential equation (\ref{dglstructure}) can be solved by 
the variation of constant method. To apply this we first determine 
the solution of the homogeneous equation. The scaling can be extracted 
if only the leading singular contribution (for $\Lambda \to
0$) of ${\mathcal F}^{(2)}_{k',k}(\Lambda, \Sigma^\Lambda)$ Eq.~(\ref{F2eval}) 
at $k'=k_F$ and $k=-k_F$ is kept. Following the same steps as in
Ref.~\onlinecite{VM1} we find 
\begin{eqnarray}
 \frac{\partial}{\partial\Lambda}
 \left[\Sigma^{\Lambda}_{k_F,-k_F}\right]_{\rm hom} 
\approx - \frac{U}{2 \pi \tau} \; \frac{1}{\Lambda}
 \left[ \Sigma^{\Lambda}_{k_F,-k_F}\right]_{\rm hom}   \; ,
\label{dglhom}
\end{eqnarray}
with the solution 
\begin{eqnarray}
 \left[\Sigma^{\Lambda}_{k_F,-k_F}\right]_{\rm hom}/\tau \approx 
c_0 \;  (\Lambda/\tau)^{- U/(2 \pi \tau)} \, ,
\label{dglhomsol}
\end{eqnarray}
where $c_0$ is a dimensionless constant. The singular part 
of the solution of the inhomogeneous linear
differential equation is then given by 
\begin{eqnarray}
\Sigma^{\Lambda}_{k_F,-k_F}/\tau \approx 
c(\Lambda) \;  (\Lambda/\tau)^{- U/(2 \pi \tau)} \, ,
\label{dglsol}
\end{eqnarray}
with
\begin{eqnarray}
c(\Lambda) = \int_{\Lambda_0}^{\Lambda} d \Lambda' \;
\frac{1}{\left[\Sigma^{\Lambda'}_{k_F,-k_F}\right]_{\rm hom}}  
\; {\mathcal F}^{(1)}_{k_F,-k_F}(\Lambda') \; .
\label{fLambdadef}
\end{eqnarray}
For small $\Lambda$ the function $c(\Lambda)$ is non-singular and
given by 
\begin{eqnarray}
\lim_{\Lambda \to 0} c(\Lambda) =  -\frac{U}{\tau} \tilde h(-2k_F) e^{-i
  k_F} \; \bar c
\label{limf}
\end{eqnarray}
with $\bar c$ being a constant of order $1$. This gives 
\begin{eqnarray}
\Sigma^{\Lambda}_{k_F,-k_F}/\tau \approx 
-\frac{U}{\tau} \tilde h(-2k_F) e^{-i
  k_F} \; \bar  c \;  (\Lambda/\tau)^{- U/(2 \pi \tau)} 
\label{dglsolcomplete}
\end{eqnarray}
and with the Born approximation the final result
\begin{eqnarray}
\frac{1}{2 \pi} - G \sim  \left| U/\tau \right|^2 \; 
\left|  \tilde h(-2k_F) \right|^2 \; (\Lambda/\tau)^{- U/(\pi \tau)} \; .
\label{thatsit}
\end{eqnarray}

To obtain the singular part of the solution of the homogeneous
differential equation $ \left[\Sigma^{\Lambda}_{k_F,-k_F}\right]_{\rm
  hom}$ we assumed that the interaction is turned on and off
abruptly. One can show that the parts of $h(j)$ with a smooth
variation of finite length do not contribute to the singular 
part of Eq.~(\ref{F_2def}). Thus  the same singular part is found 
independently of how the interaction is varied and 
Eq.~(\ref{thatsit}) is valid for general $h(j)$. 

The power-law scaling Eq.~(\ref{thatsit}) 
holds for $\Lambda_c \ll \Lambda \ll \tau$ 
with a scale $\Lambda_c$ set by 
\begin{eqnarray}
\left| U/\tau \right|^2 \; 
\left|  \tilde h(-2k_F) \right|^2 \; (\Lambda_c/\tau)^{- U/(\pi \tau)}
\sim 1/(4 \pi)\; .
\label{thatsitscale}
\end{eqnarray} 
To leading order in $U/\tau$ the exponent $U/(\pi \tau)$ agrees with the 
exponent $2(K-1)/(K+1)$ [see Eq.~(\ref{leadingorder})] 
found for a single weak impurity placed close to a perfect 
(that is arbitrarily smooth) contact.\cite{FurusakiNagaosa,Tilman} 
We thus have analytically shown, that with respect to the 
scaling exponent  weak single-particle and weak two-particle   
inhomogeneities are indeed equivalent. The leading order 
exponent is furthermore consistent with the numerical results of 
the last section. 

For the single impurity case the prefactor of the power law is given
by $ \left|V_{\rm back}/\tau\right|^2$. In case of the two-particle
inhomogeneity this is replaced by the square of the bulk interaction
and the square of the $2k_F$ Fourier component (backscattering) of the
function $h(j)$ with which the interaction is turned on and
off. The smaller this component the smaller is the perturbation due to
the two contacts. 
Therefore the smoothness of the turning on and off is directly
measured by $\tilde h(-2 k_F)$. 
The presence of the factor 
$|U/\tau|^2$ explains why in the numerical study the weak inhomogeneity 
exponent for larger $U/\tau$ is only observable for fairly smooth
contacts, that is $h(j)$'s with small $2 k_F$ component. For larger
$U/\tau$ and fairly abrupt contacts the rhs of Eq.~(\ref{thatsit})
becomes too large already on intermediate energy scales and no 
energy window for scaling is left. For $ \Lambda \ll \Lambda_c$ 
defined in Eq.~(\ref{thatsitscale}) the inhomogeneity is effectively 
large and Eq.~(\ref{vanish}) holds.

We here considered the half-filled band case, but also other fillings
can be studied following the same steps.\cite{VM1}

The analysis Eqs.~(\ref{flowsigmamomentum}) to (\ref{thatsit}) also 
holds if in addition weak single-particle impurities are placed close 
to the contacts,
with the only difference that $\Sigma^{\Lambda_0}_{k_F,-k_F}$ now has a
non-vanishing initial condition set by the backscattering of the 
bare impurity. The prefactor in Eq.~(\ref{thatsit}) is determined 
by either the square of the single impurity backscattering amplitude or
$ \left| U/\tau \right|^2 \; \left|  h(-2k_F) \right|^2$ depending on
the relative size.

\section{Summary}
\label{summary}

In the present paper we have investigated the role of contacts,
defined by an inhomogeneous two-particle interaction, on the linear
conductance through an interacting 1d quantum wire. 
The wire and contacts were assumed to be free of any bare single-particle
impurities. We first showed that within the LLL picture the contacts
are always perfect, that is the conductance is $1/(2 \pi)$ independent
of the strength of the interaction, the length of the wire, and the
spatial variation of the LL parameters. Earlier only step-wise changes
of the LL parameters were considered. We then studied the problem
within a microscopic lattice model. 
Similar to the case of a single impurity the interplay of the two-particle 
inhomogeneity and the bulk interaction generates an oscillating 
self-energy with an amplitude that decays slowly away from the 
contacts. Scattering off this effective potential leads to the 
discussed effects. We showed that within the microscopic 
modeling $1 /(2 \pi)- G$ increases with increasing bulk interaction $U$, 
increasing wire length $N$, and decreasing smoothness. The 
measure for smoothness is given by $\tilde h(-2k_F)$, that is
the backscattering Fourier component of the spatial variation $h(j)$ of the
two-particle interaction. As long as the inhomogeneity stays
effectively small $1/(2 \pi)-G$ shows power-law scaling with an
exponent that is consistent with $2(K-1)/(K+1)$, the scaling exponent
known for the case of a wire with a single impurity placed close to
one of the two smooth contacts. 

In experiments on quantum wires the leads are electronically two- or
three-dimensional. Depending on the systems studied the contacts are either
regions in which the system gradually crosses over from higher
dimensions to quasi 1d or the contact regions extend over a finite
part of the wire (see the experiments on carbon
nanotubes).\cite{experiments} This shows that our simplified
description---1d leads, spatially dependent interactions close to end
contacts, no explicit single-particle scattering at the
contacts---provides only an additional step towards a detailed 
understanding of the role of contacts in transport through interacting 
quantum wires, such as carbon nanotubes and cleaved edge overgrowth 
samples.\cite{experiments}

\section*{Acknowledgments}
We thank S.~Jakobs and H.~Schoeller for very fruitful discussions. 
The authors are grateful to the Deutsche Forschungsgemeinschaft 
(SFB 602) for support.


\begin{thebibliography}{99}
\bibitem{comment} There is no generally accepted use of prefactors
in the definition of the $\Phi_\alpha$
in the literature. We therefore did not include any.
\bibitem{footnote1} 
Under quite weak assumptions on the frequency
dependence of the self-energy, that are fulfilled for the present 
situation, there are no current vertex corrections at $T=0$ and
Eq.~(\ref{conductf}) is exact (Ref.~\onlinecite{Oguri2}).
\end{thebibliography}
\end{document}